\documentclass[a4paper,11pt]{article}

\usepackage{apalike}
\usepackage{graphicx}

\newcommand{\smean}[1]{\langle #1 \rangle}

\begin{document}

\title{Spatiotemporal adaptation through corticothalamic loops: A hypothesis}
\author{Ulrich Hillenbrand\thanks{Present address: Institute of Robotics and Mechatronics, German Aerospace Center, Oberpfaffenhofen, 82234 Wessling, Germany. Email: Ulrich.Hillenbrand@dlr.de} and J.\ Leo van Hemmen\\Physik Department, TU M\"unchen\\85747 Garching bei M\"unchen\\Germany}

\date{}

\maketitle

\begin{abstract}
The thalamus is the major gate to the cortex and its control over cortical
responses is well established. Cortical feedback to the thalamus is, in turn,
the anatomically dominant input to relay cells, yet its influence on thalamic
processing has been difficult to interpret. For an understanding of complex
sensory processing, detailed concepts of the corticothalamic interplay need yet
to be established. Drawing on various physiological and anatomical data, we
elaborate the novel hypothesis that the visual cortex controls the
spatiotemporal structure of cortical receptive fields via feedback to the
lateral geniculate nucleus. Furthermore, we present and analyze a model of
corticogeniculate loops that implements this control, and exhibit its ability
of object segmentation by statistical motion analysis in the visual field.
\end{abstract}

\ \\
\noindent{\bf Keywords:} Lateral geniculate nucleus, Corticothalamic feedback,
Control loop, Object segmentation, Motion analysis.\\
\ \\
\noindent Published as {\em Visual Neuroscience} {\bf 17} (2000), pp.\ 107--118.
\ \\

\section*{Introduction}

It has long been an intriguing problem that cortical feedback to the thalamus
is the anatomically dominant input to relay cells, yet the modulation it
effects in their responses seems to be ambiguous. Experiments and theoretical
considerations have suggested a variety of operations of the visual cortex on
the lateral geniculate nucleus (LGN), such as attention-related gating of
geniculate relay cells (GRCs) \cite{Sherman&Koch86,Koch87}, synchronizing
firing of GRCs \cite{Sillito_etal94,Singer94}, increasing mutual information
between GRCs' retinal input and their output \cite{McClurkin_etal94}, and
switching GRCs from a detection to an analyzing mode
\cite{Sherman&Guillery96,Godwin_etal96}. Most of the current speculations,
however, fall short of a clear, integrated view of corticothalamic function.
Detailed concepts of the relationship between thalamic and cortical operation
still remain to be established so as to advance our ideas about complex
sensory, and ultimately cognitive, processing. In the present article we
provide support for a role of corticogeniculate loops in complex visual motion
processing.

In the A laminae of cat LGN two types of X relay cell have been identified that
dramatically differ in their temporal response properties
\cite{Mastronarde87a,Humphrey&Weller88a,Saul&Humphrey90}. Those that are more
delayed in response time and phase have been termed {\em lagged}, the others
{\em nonlagged} cells (with the exception of very few so-called {\em partially
lagged} neurons). In particular, the peak response of lagged neurons to a
moving bar is about 100 ms later than that of nonlagged neurons
\cite{Mastronarde87a}. Lagged X cells comprise about 40 \% of all X relay cells
\cite{Mastronarde87a,Humphrey&Weller88b}. Physiological \cite{Mastronarde87b},
pharmacological \cite{Heggelund&Hartveit90}, and structural
\cite{Humphrey&Weller88b} evidence suggests that rapid feedforward inhibition
via intrageniculate interneurons plays a decisive role in shaping the lagged
cells' response. Some authors have additionally related differences in receptor
types to the lagged-nonlagged dichotomy
\cite{Heggelund&Hartveit90,Hartveit&Heggelund90}; see, however, Kwon et al.\
(1991)\nocite{Kwon_etal91}.

Layer 4B in cortical area 17 of the cat is the target of both lagged and
nonlagged geniculate X cells
\cite{Saul&Humphrey92a,Jagadeesh_etal97,Murthy_etal98}. The spatiotemporal
receptive fields (RFs) of its direction-selective simple cells can routinely be
interpreted as being composed of subregions that receive geniculate inputs
alternating between lagged and nonlagged X type
\cite{Saul&Humphrey92a,Saul&Humphrey92b,DeAngelis_etal95,Jagadeesh_etal97,Murthy_etal98}.
At least for simple cells in layer 4B, this RF structure determines the
response to moving visual features
\cite{McLean&Palmer89,Reid_etal91,DeAngelis_etal95,Jagadeesh_etal93,Jagadeesh_etal97,Murthy_etal98},
and thus the cell's tuning for direction and speed\footnote{To avoid confusion,
we point out that the term `speed tuning' is sometimes used in a more
restricted sense. Simple cells exhibit tuning for spatial and temporal
frequencies that results in preference for speeds of moving grids depending on
their spatial frequency. Here we will be concerned with the more natural case
of stimuli having a low-pass frequency content \cite{Field94}, specifically,
those composed of local features such as thin bars.\label{footnote1}}.

Certainly, intracortical input to cortical cells contributes to
direction-selective responses, given that these inputs anatomically outnumber
thalamic inputs \cite{Ahmed_etal94}. Suggested intracortical effects include
sharpening of tuning properties by suppressive interactions
\cite{Reid_etal91,Hirsch_etal98,Crook_etal98} and
amplification of geniculate inputs by recurrent excitation
\cite{Douglas_etal95,Suarez_etal95}. Intracortical circuits can in principle
even generate their own direction selectivity by selectively inhibiting
responses to nonpreferred motion
\cite{Douglas_etal95,Suarez_etal95,Maex&Orban96}. Our modeling is complementary
to the latter in that we emphasize the influence of geniculate inputs on
cortical RF properties that is suggested by numerous studies
\cite{Saul&Humphrey92a,Saul&Humphrey92b,Reid&Alonso95,Alonso_etal96,Ferster_etal96,Jagadeesh_etal97,Murthy_etal98,Hirsch_etal98},
in order to bring out effects that are specific to the geniculate contribution
to spatiotemporal tuning.

Thalamocortical neurons possess a characteristic blend of voltage-gated ion
channels that jointly determine the timing and pattern of Na$^+$ spiking in
response to a sensory stimulus.  Depending on the initial membrane
polarization, the GRC response to a visual stimulus is in a range between a
{\em tonic} and a {\em burst} mode \cite{Sherman&Guillery96}.  At
hyperpolarization, a Ca$^{2+}$ conductance gets de-inactivated and, on
activation, promotes burst firing. Although the issue is still controversial,
there is evidence that a {\em mixture} of burst and tonic spikes is involved in
the transmission of visual signals in lightly anesthetized or awake animals
\cite{Guido_etal92,Guido_etal95,Guido&Weyand95,Mukherjee&Kaplan95,Sherman&Guillery96,Reinagel_etal99}.
In nonlagged neurons a burst component is present at resting membrane
potentials below roughly -70 mV and constitutes a very early part of a visual
response \cite{Lu_etal92,Guido_etal92,Mukherjee&Kaplan95}. In lagged neurons
bursting seems to be responsible for high-activity transients seen after the
offset of feedforward inhibition; in particular, it is contributing
substantially to the delayed peak response to a moving bar
\cite{Mastronarde87b}.

Cortical feedback to the A laminae of the LGN, arising mainly from layer 6 of
area 17 \cite{Sherman&Guillery96}, can locally modulate the response mode, and
hence the {\em timing}, of GRCs by shifting their membrane potentials on a time
scale that is long as compared to retinal inputs.  This may occur directly
through the action of metabotropic glutamate and NMDA receptors
(depolarization) \cite{McCormick&vonKrosigk92,Godwin_etal96,Sherman&Guillery96}
and indirectly via the perigeniculate nucleus (PGN) or geniculate interneurons
by activation of GABA$_{\rm B}$ receptors (hyperpolarization) of GRCs
\cite{Crunelli&Leresche91,Sherman&Guillery96}. Indeed, GRCs in vivo are dynamic
and differ individually in their degree of burstiness
\cite{Lu_etal92,Guido_etal92,Mukherjee&Kaplan95}. Here we explicate the causal
link between the {\em variable response timing of GRCs} and {\em variable
tuning of cortical simple cells for speed} of moving features, thus identifying
control of speed tuning as a likely mode of corticothalamic operation.
Moreover, we exemplify the computational power of the hypothesized control
mechanism in a model of corticogeniculate loops that performs object
segmentation based on motion cues.

\section*{Geniculate input to the cortex}

The first object of our study is the geniculate input to a cortical neuron in
layer 4B.

\subsection*{Model of the primary visual pathway}

\begin{figure}
\begin{minipage}{0.5\textwidth}
\leftline{\includegraphics[width=0.9\textwidth]{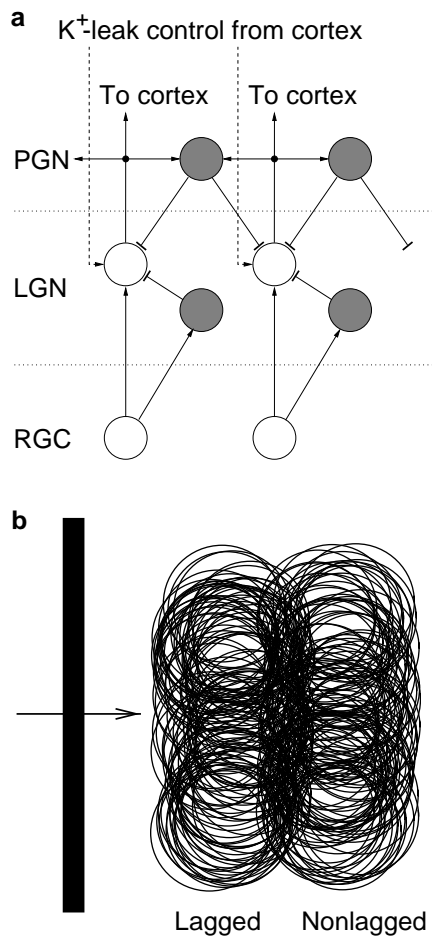}}
\end{minipage}
\begin{minipage}{0.49\textwidth}
\caption[]{Model of the primary visual pathway. ({\bf a}) Sketch of the lateral
geniculate nucleus (LGN). Open/filled circles and arrow/bar heads indicate
excitatory/inhibitory neurons and their respective synapses. A retinal ganglion
cell (RGC) synapses excitatorily on a geniculate relay cell (GRC) and on an
intrageniculate interneuron, which in turn inhibits the same GRC (arrangement
called `synaptic triad'). The relative strengths of feedforward excitation and
feedforward inhibition shape a GRC's response to be of the lagged or nonlagged
type; see main text and Fig.\ \ref{fig2}. There is an inhibitory feedback loop
via the perigeniculate nucleus (PGN). The influence of cortical feedback has
been modeled as a variation of the GRCs' resting membrane potential by control
of a K$^+$ leak current. ({\bf b}) Arrangement in visual space of the receptive
field (RF) centers of the 100 lagged and 100 nonlagged GRCs comprising the
model LGN. These GRCs are envisaged to project onto the same cortical simple
cell and create an on or off region of its RF. In the simulations, the diameter
of a single lagged or nonlagged RF center is 0.5 degrees. Results for rescaled
versions of this geometry can be derived straightforwardly from the simulations;
see main text. The bar and arrow on the left indicate preferred orientation
and direction of motion, respectively.}
\label{fig1}
\end{minipage}
\end{figure}

For the GRCs we have employed a 12-channel model of the cat relay neuron
\cite{Huguenard&McCormick92,McCormick&Huguenard92}, adapted to 37 degrees
Celsius. It includes a transient and a persistent Na$^+$ current, several
voltage-gated K$^+$ currents, a voltage- and Ca$^{2+}$-gated K$^+$ current, a
low- and a high-threshold Ca$^{2+}$ current, a hyperpolarization-activated
mixed cation current, and Na$^+$ and K$^+$ leak conductances. As shown in Fig.\
\ref{fig1}{\bf a}, retinal input reaches a GRC directly as excitation, and
indirectly via an intrageniculate interneuron as inhibition, thus establishing
the typical triadic synaptic circuit found in the glomeruli of X GRCs
\cite{Sherman&Guillery96}. The temporal difference between the two afferent
pathways equals the delay of the inhibitory synapse and has been taken to be
1.0 ms \cite{Mastronarde87b}.

It is known that both NMDA and non-NMDA receptors contribute to
retinogeniculate excitation to varying degrees, ranging from almost pure
non-NMDA to almost pure NMDA mediated responses in individual GRCs of both
lagged and nonlagged varieties \cite{Kwon_etal91}. At least in lagged cells,
however, early responses and, hence, responses to the transient stimuli that
will be considered here, seem to depend to a lesser degree on the NMDA receptor
type than late responses \cite{Kwon_etal91}. Since the essential
characteristics of lagged and nonlagged responses apparently do not depend on
the special properties of NMDA receptors -- an assumption confirmed by our
results -- we chose the postsynaptic conductances in GRCs to be of the
non-NMDA type.

The time course of postsynaptic conductance change in GRCs following reception
of an input has been modeled by an alpha function,
\begin{equation}
{\rm g}(t>0) \, = \, g_{\rm syn} \, \frac{t}{\tau_{\rm syn}} \,
\exp\!\left(1 - \frac{t}{\tau_{\rm syn}}\right) \ .
\label{g(t)}
\end{equation}
For excitation, the rise time $\tau_{\rm syn}$ has been chosen to be 0.4 ms
\cite{Mukherjee&Kaplan95}, for inhibition it is 0.8 ms. The latter value was
estimated from the relative durations of S potentials recorded at excitatory
and inhibitory geniculate synapses \cite{Mastronarde87b} and was found to
reproduce the rise times of inhibitory postsynaptic potentials recorded in
relay cells following stimulation of the optic chiasm
\cite{Bloomfield&Sherman88}. The reversal potentials are for excitation 0 mV,
for inhibition $-$85.8 mV \cite{Bal_etal95a}.

\begin{figure}
\centerline{\includegraphics[width=0.6\textwidth]{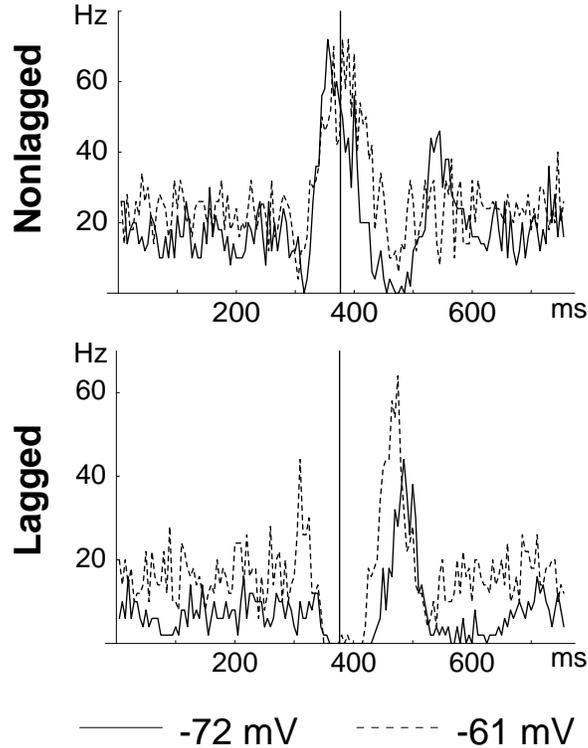}}
\caption[]{Numerical simulation of moving-bar response of single relay neurons.
Typical nonlagged (top row) and lagged (bottom row) responses to a bar moving
at 4 deg/s (100 sweeps averaged, bin width 5 ms) have been reproduced by relay
neurons in the LGN model (Fig.\ \ref{fig1}{\bf a}) at the two resting membrane
potentials $-$61 mV (dashed line) and $-$72 mV (solid line). The vertical line
in the center of the plots indicates the time of the retinal input peak. As the
membrane is {\em hyperpolarized}, the nonlagged bar-response peak shifts to
{\em earlier} times. Conversely, the lagged response shifts to {\em later}
times. The nonlagged responses are produced by strong feedforward excitation
[$g_{\rm syn} = 0.05 \mu$S; cf.\ eqn.\ (\ref{g(t)})] and weak feedforward
inhibition ($g_{\rm syn} = 0.0125 \mu$S), whereas the lagged responses
correspond to weak feedforward excitation ($g_{\rm syn} = 0.0125 \mu$S) and
strong feedforward inhibition ($g_{\rm syn} = 0.25 \mu$S). The numerical values
of $g_{\rm syn}$ are not critical. The response timing is consistent with
experiments \cite{Mastronarde87a,Lu_etal92,Guido_etal92}. Furthermore, we have
found the temporal transfer functions (not shown) to agree well with published
data \cite{Saul&Humphrey90,Mukherjee&Kaplan95}.}
\label{fig2}
\end{figure}

We have found typical lagged responses for strong feedforward inhibition with
weak feedforward excitation, in agreement with Mastronarde (1987b), Humphrey \&
Weller (1988b), and Heggelund \& Hartveit
(1990)\nocite{Mastronarde87b,Humphrey&Weller88b,Heggelund&Hartveit90}. On the
other hand, typical nonlagged responses are produced by weak feedforward 
inhibition with strong feedforward excitation; see Fig.\ \ref{fig2}. We have
therefore implemented lagged and nonlagged relay cells in the model by varying
the relative strengths of feedforward excitation and feedforward inhibition.
For an explanation of lagged and nonlagged responses see the results below.

The model system comprises 100 lagged and 100 nonlagged relay neurons. Their RF
centers are 0.5 degrees in diameter \cite{Cleland_etal79} and are spatially
arranged in a lagged and a nonlagged clusters subtending 0.7 degrees each and
displaced by 0.45 degrees; see Fig.\ \ref{fig1}{\bf b}. This layout matches the
basic structure of an on or off region in a RF of a directional simple cell in
cortical layer 4B onto which the GRCs are envisaged to project
\cite{Saul&Humphrey92a,Saul&Humphrey92b,DeAngelis_etal95,Jagadeesh_etal97,Murthy_etal98}.
To complete the geniculate input to a RF of this type, this lagged-nonlagged
unit would have to be repeated with alternating on-off polarity and a spatial
offset that would determine the simple cell's preference for some spatial
frequency. Since we are not concerned here with effects of spatial frequency
(see previous footnote \ref{footnote1}), omission of the other on/off regions
does not affect our conclusions. Results for rescaled RF geometries can be
derived straightforwardly; see the results below. The number of geniculate
cells contributing to the simple cell's RF has been estimated roughly from
Ahmed et al.\ (1994)\nocite{Ahmed_etal94}. Only its order of magnitude matters.

We have also taken into account feedback inhibition via the PGN
\cite{Lo&Sherman94,Sherman&Guillery96}; see Fig.\ \ref{fig1}{\bf a}.
Connections between PGN neurons and GRCs are all to all within, and separate
for the lagged and nonlagged populations. Axonal plus synaptic delays are 2.0
ms in both directions.

Intrageniculate interneurons and PGN cells, like GRCs, posses a complex blend
of ionic currents. They are, however, thought to be active mainly in a tonic
spiking mode during the awake state \cite{Contreras_etal93,Pape_etal94}. For an
efficient usage of computational resources and time we have therefore modeled
these neurons by the spike-response model \cite{Gerstner&vanHemmen92}, which
gives a reasonable approximation to tonic spiking \cite{Kistler_etal97}. Note
that for the present model it is irrelevant whether transmission across
dendrodendritic synapses between intrageniculate interneurons and GRCs actually
occurs with or without spikes; cf.\ Cox et al.\ (1998)\nocite{Cox_etal98}. For
a relay neuron, all that matters is the fact that an excitatory retinal input
is mostly followed by an inhibitory input \cite{Bloomfield&Sherman88}. The
spike-response neurons have been given an adaptive spike output
\cite{Gerstner&vanHemmen92}, i.e., there is adaptation of transmission
across the dendrodendritic synapses.

We have incorporated the influence of cortical feedback to the thalamus by
varying the K$^+$ leak conductivity of GRCs, which in turn controls their
resting membrane potential \cite{McCormick&vonKrosigk92,Godwin_etal96}; see
Fig.\ \ref{fig1}{\bf a}. All GRCs, lagged and nonlagged, have been assigned the
same resting membrane potential; here we assume a uniform action of cortical
feedback on the scale of single RFs in area 17. By varying the resting membrane
potential we investigate a strictly modulatory role of corticogeniculate
feedback, as opposed to the retinal inputs that drive relay cells to fire; cf.\
Sherman \& Guillery (1996), Crick \& Koch
(1998)\nocite{Sherman&Guillery96,Crick&Koch98}.

The model is described by a high-dimensional system of nonlinear, coupled
differential equations \cite{Huguenard&McCormick92,McCormick&Huguenard92}. The
input to GRCs was modeled as a set of Poisson spike trains with time-varying
firing rates as recorded from retinal ganglion cells in response to moving,
thin (0.1 degrees), long (10 degrees) bars \cite{Cleland&Harding83}. For
numerical integration of the stochastic differential equations we used an
adaptive fifth-order Runge-Kutta algorithm. The dynamics of spike-response
neurons has been solved by exact integrals. Simulations were run on an IBM SP2
parallel computer.

We collected spike times with 0.1 ms resolution. For each velocity $v$ of bar
motion tested, spikes of all 100 lagged/nonlagged relay cells were pooled. We
calculated the total lagged/nonlagged response rates ${\rm r}_{\rm
l}(v,t)$/${\rm r}_{\rm nl}(v,t)$ as spike counts in 5 ms windows, a timescale
relevant to postsynaptic integration, shifted by steps of 1 ms, i.e.,
$t=1,2,\ldots$ ms. The velocity tuning of the pooled lagged/nonlagged peak rate
per neuron is
\begin{equation}
{\rm R}_{\ell}(v) \, = \, \frac{1}{100} \, \max_{t \in [t_{\rm i},t_{\rm f}]}
{\rm r}_{\ell}(v,t) \ ,
\quad \ell = {\rm l},{\rm nl} \ ,
\end{equation}
where the times $t_{\rm i}$ and $t_{\rm f}$ are chosen such that all responses
lie in the interval $[t_{\rm i},t_{\rm f}]$.

We are primarily interested in the {\em total geniculate input} to a cortical
simple cell. To this end, we have shifted lagged spikes by 2 ms to later times
in order to account for the fact that the lagged cells' conduction times to
cortex are slightly longer than those of the nonlagged cells
\cite{Mastronarde87a,Humphrey&Weller88a}. Furthermore, although lagged
responses in the LGN tend to be weaker than nonlagged responses
\cite{Mastronarde87a,Humphrey&Weller88a,Saul&Humphrey90}, they appear to be
about equally effective in driving cortical simple cells
\cite{Saul&Humphrey92a}. The cortical (possibly synaptic) cause being beyond
the scope of this work, we simply have counted every lagged spike twice to
obtain the velocity tuning of the effective geniculate input to a cortical cell,
\begin{equation}
{\rm R}(v) \, = \, \frac{1}{100} \, \max_{t \in [t_{\rm i},t_{\rm f}]} \left[
2 \, {\rm r}_{\rm l}(v,t - 2 \, {\rm ms}) + {\rm r}_{\rm nl}(v,t) \right] \ .
\end{equation}
The peak input rate per lagged-nonlagged pair ${\rm R}(v)$ is correlated with
simple-cell activity because postsynaptic potentials are summed almost linearly
in simple cells \cite{Jagadeesh_etal93,Kontsevich95,Jagadeesh_etal97}.

The total geniculate input rate to a cortical neuron depends on (i) the
magnitude of the pooled lagged and nonlagged response peaks, ${\rm R}_{\rm
l}(v)$ and ${\rm R}_{\rm nl}(v)$, respectively, and (ii) their relative timing.
To differentiate between these two factors we have determined the times ${\rm
t}_{\rm l}(v)$ and ${\rm t}_{\rm nl}(v)$ of the maxima of the lagged and
nonlagged response rates, respectively,
\begin{equation}
{\rm t}_{\ell}(v) \, = \, {\rm arg} \max_{t \in [t_{\rm i},t_{\rm f}]}
{\rm r}_{\ell}(v,t) \ ,
\quad \ell = {\rm l},{\rm nl} \ ,
\end{equation}
and calculated the peak-time differences ${\rm t}_{\rm nl}(v) - {\rm t}_{\rm
l}(v)$ as a function of the bar velocity. Means and standard errors have been
estimated from a sample of 30 sweeps at each bar velocity.

\subsection*{Results for geniculate input to the cortex}

\begin{figure}
\centerline{\includegraphics[width=\textwidth]{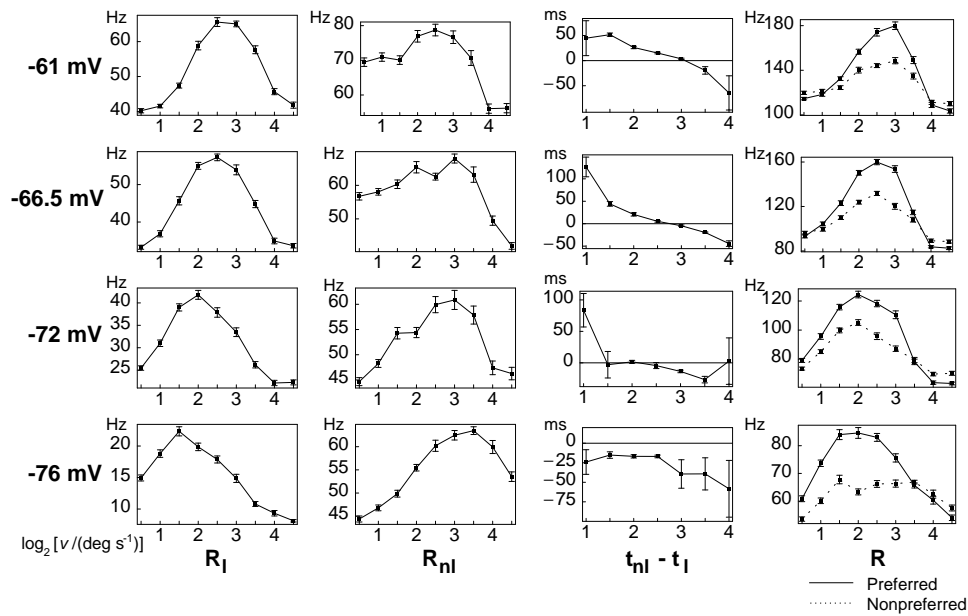}}
\caption[]{Numerical simulation of geniculate moving-bar response and input to
cortex. Velocity tuning and timing of the response peaks have been plotted for
resting membrane potentials indicated on the far left. In the columns we show
from left to right as functions of the bar velocity the peak response rate of
the lagged population (${\rm R}_{\rm l}$), of the nonlagged population (${\rm
R}_{\rm nl}$), their peak-time difference (${\rm t}_{\rm nl} - {\rm t}_{\rm
l}$) for the preferred direction, and the peak of the total geniculate input
(R) to a cortical cell for the preferred (solid line) and nonpreferred (dotted
line) direction of motion. The horizontal axes show the logarithm (base 2) of
speed in all graphs. The bars in the graphs are standard errors. As the
membrane is {\em hyperpolarized}, the total geniculate input to a cortical
cell, and hence that cortical cell, prefer progressively {\em lower}
velocities. Means and standard errors are estimated from 30 bar sweeps.}
\label{fig3}
\end{figure}

For different values of the resting membrane potential, Fig.\ \ref{fig3} shows
in the columns from left to right the velocity tuning of the lagged population
(${\rm R}_{\rm l}$), of the nonlagged population (${\rm R}_{\rm nl}$), the
peak-time differences (${\rm t}_{\rm nl} - {\rm t}_{\rm l}$) of their responses
for the preferred direction, and the tuning of the total geniculate input (R)
to a cortical cell for the preferred and nonpreferred direction of motion; see
above for details. As in vivo, the lagged cells prefer lower velocities and
have lower peak firing rates than the nonlagged cells
\cite{Mastronarde87a,Humphrey&Weller88a,Saul&Humphrey90}. The key observation,
however, is that the maximum of the {\em total geniculate input rate} to a
cortical neuron shifts to {\em lower velocities} as the membrane potential {\em
hyperpolarizes}; see Fig.\ \ref{fig3} right column.

The total geniculate input rate R assumes its maximum at a velocity of bar
motion where the peak discharges of the lagged and nonlagged neurons {\em
coincide}, i.e., where ${\rm t}_{\rm nl} - {\rm t}_{\rm l} \approx 0$. The {\em
shift} of the maximum with hyperpolarization to lower velocities is produced by
a corresponding shift of the peak-time differences ${\rm t}_{\rm nl} - {\rm
t}_{\rm l}$ and of the lagged tuning ${\rm R}_{\rm l}$, while the peak of the
nonlagged tuning ${\rm R}_{\rm nl}$ remains essentially unchanged. As is
demonstrated in Fig.\ \ref{fig2}, the change in peak-time difference is due to
(i) a shift of the lagged response peak to {\em later} times, and (ii) a shift
of the nonlagged response peak to {\em earlier} times.

The reason for the opposite shifts of lagged and nonlagged response timing lies
in the interaction of the low-threshold Ca$^{2+}$ current with the different
levels of inhibition received by lagged and nonlagged neurons. With only weak
feedforward inhibition, nonlagged neurons respond to retinal input with
immediate depolarization, eventually reaching the activation threshold for the
Ca$^{2+}$ current. If the Ca$^{2+}$ current is in the de-inactivated state, it
will boost depolarization and give rise to an early burst component of the
visual response \cite{Lu_etal92,Guido_etal92}. The lower the resting membrane
potential, the more de-inactivated the Ca$^{2+}$ current will be, and the
stronger the early response component relative to the late tonic component. In
an ensemble of neurons that receive retinal input at slightly different times,
like the 100 nonlagged neurons with spatially scattered RFs (cf.\ Fig.\
\ref{fig1}{\bf b}), this leads to a {\em gradual} shift of the
ensemble-response maximum with membrane polarization. Lagged neurons, on the
other hand, receive strong feedforward inhibition and, hence, initially respond
to retinal input with hyperpolarization. Repolarization occurs when inhibition
gets weaker due to declining retinal input rate or adaptation of the inhibitory
input. With the Ca$^{2+}$ current being de-inactivated by the excursion of the
membrane potential to low values, lagged spiking starts with burst spikes as
soon as the voltage reaches the Ca$^{2+}$-activation threshold. This will take
longer, if the resting membrane potential is lower, leading to the observed
shift in response timing with membrane polarization.

Not surprisingly, the total geniculate input rate R is higher for the direction
of bar motion where ${\rm t}_{\rm nl} - {\rm t}_{\rm l}$ assumes lower values.
In other words, the direction preferred is the one where the lagged cells
receive their retinal input before the nonlagged cells; cf.\ Figs.\
\ref{fig1}{\bf b} and \ref{fig3} right column.

We have investigated  the geniculate input to simple cells, which clearly
cannot be compared with their output directly. Nonetheless it is interesting to
note that, much like velocity tuning in areas 17 and 18 \cite{Orban_etal81a},
the modeled geniculate input at the optimal velocity decreases and the tuning
width increases with decreasing optimal velocity; see Fig.\ \ref{fig3} right
column.

Because of scaling properties of the retinal ganglion cells' velocity tuning
\cite{Cleland&Harding83}, rescaled versions of the RF geometry shown in Fig.\
\ref{fig1}{\bf b} produce accordingly shifted tuning curves (on a logarithmic
speed scale). In particular, we retrieve the positive correlation between RF
size and preferred speed found in areas 17 and 18 \cite{Orban_etal81a} from the
geniculate input.

\section*{The corticogeniculate loop}

\begin{figure}
\centerline{\includegraphics[width=\textwidth]{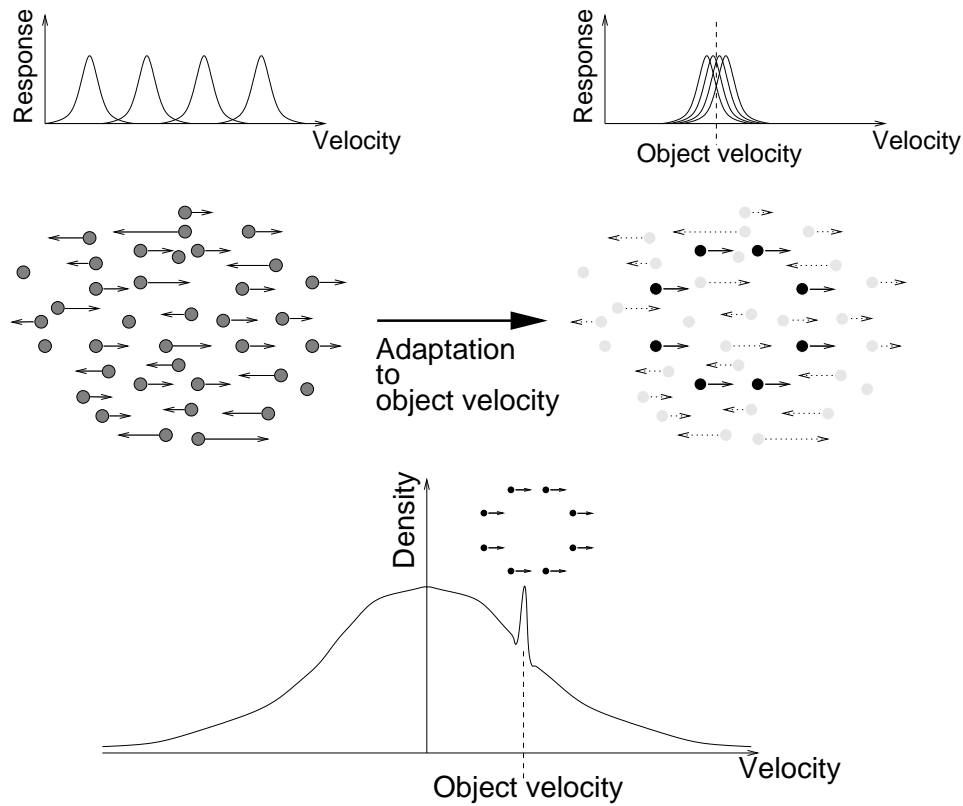}}
\caption[]{The rationale of object segmentation by adaptive velocity tuning.
The tuning curves shown at the top of the figure schematically represent the
peak response rates of (four) cortical neurons as a function of the speed of a
local feature passing their receptive field in an unadapted state (top left)
and an adapted state (top right). Within some region of the visual field a
natural stimulus consists of a collection of local features (depicted as dots,
center) moving from left to right and from right to left at various speeds
(depicted as arrows). A subset of them is moving at a {\em common} speed from
left to right. The velocity density (bottom) of this type of stimulus consists
of two components: one symmetrical with respect to the two directions of motion
and one asymmetrical. The former derives from the incoherent, the latter from
the coherent motion and is the statistical signature of a moving object. The
adaptive motion system has to detect and tune in to the asymmetrical component
of the velocity density. After adaptation of cortical velocity tuning (top
right), object features are prominent in cortical representation, whereas other
features are suppressed (center right). The given stimulus scenario generalizes
straightforwardly to motion in two dimensions.}
\label{vadapt_osegm}
\end{figure}

Once a dynamic gating mechanism for thalamocortical information transfer has
been identified, one has to face the key question of how gating can be
controlled. The computational goal of this control should certainly be an
enhanced cortical representation of behaviorally relevant stimuli, generally
referred to as {\em objects}, and suppression of less significant bits, such as
neuronal noise and incoherent background motion. The general idea of object
segmentation by adaptive velocity tuning is illustrated in Fig.\
\ref{vadapt_osegm}. Since many details of corticothalamic circuits are not
known yet and since we aim at a thorough analytical treatment of the
closed-loop system for general stimulus statistics we have kept the modeling at
this point at a more abstract level than in the above case. Moreover, the
underlying principles are best exposed by a {\em simple} model.

\subsection*{Model of the corticogeniculate loop}

The membrane potential of GRCs is assumed to be modulated by feedback from
cortical layer 6. The postsynaptic potential evoked in GRCs by a burst of layer
6 action potentials fired around time $t_i$ is
\begin{equation}
{\rm PSP}(t - t_i) \, = \, \frac{t - t_i}{\tau} \, \exp\!\left(1 - \frac{t - t_i}{\tau}\right) \ ,
\quad \mbox{($t > t_i$),}
\label{PSP}
\end{equation}
that is, an alpha function. We are interested here specifically in the slow,
modulatory effect of cortical input, mediated by NMDA and metabotropic
glutamate receptors in the case of depolarization, and by GABA$_{\rm B}$
receptors for hyperpolarization. The rise time $\tau$ for combinations of NMDA
and metabotropic glutamate receptor responses and for GABA$_{\rm B}$ receptor
responses may be several 100 ms \cite{vonKrosigk_etal99}, but is kept as a
free parameter in the model, i.e., we do not specify a numerical value for
$\tau$ throughout the analysis. We neglect corticothalamic delays, which are at
least one order of magnitude smaller than $\tau$. In particular, for layer 6
projection neurons that are visually responsive and thus relevant to our
model, they are mostly below 10 ms \cite{Tsumoto_etal78,Tsumoto&Suda80}. One
can show that the inclusion of an adequate distribution of delays does not
alter the general dynamic behavior of the system, but merely adds small
corrections to some characteristic quantities.

\begin{figure}
\centerline{\includegraphics[width=\textwidth]{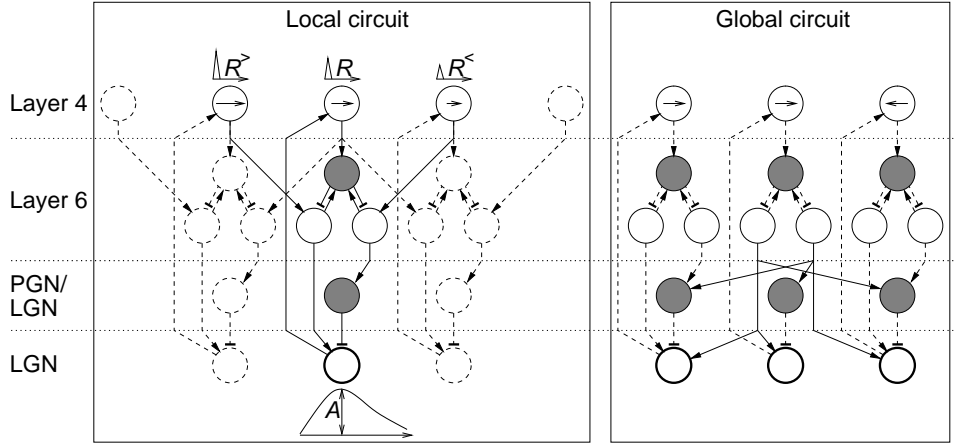}}
\caption[]{Model of corticogeniculate loop. Same conventions as in Fig.\
\ref{fig1}{\bf a}. The thick circles symbolize {\em all} geniculate relay cells
(GRCs), lagged and nonlagged, which project to a particular cortical layer 4
cell (cf.\ Fig.\ \ref{fig1}{\bf b}), as indicated by the long upward arrows. On
the left we show a local piece of circuit ({\bf Local circuit}). The cells
depicted here in layer 4 have coinciding receptive fields, the same preferred
direction but different preferred speed of motion, as indicated by the arrows
of different length (a longer arrow means a higher preferred speed). The
temporal responses of these cells to a moving feature have different rate
amplitudes (top; amplitudes $R^>$, $R$, $R^<$). Cortical feedback to the
lateral geniculate nucleus (LGN), originating from layer 6, modulates the
membrane potential of GRCs. Direct feedback depolarizes, indirect feedback via
the perigeniculate nucleus (PGN) or geniculate interneurons hyperpolarizes the
membrane of relay cells. The net effect is a slow postsynaptic potential of
amplitude $A$ in GRCs (bottom). Simulation of the local circuit shows that $A$
is well-described by eqn.\ (\ref{A}). The dashed parts of the local circuit
indicate the analogous connections that feed back to GRCs which provide input
to the other local layer 4 neurons. At the right of the figure we sketch the
global, or long-range, connectivity of the system ({\bf Global circuit}). The
local circuit, shown here only for one speed-tuned layer 4 neuron, is
replicated for different retinal positions and two opposite directions of
motion (arrows). Local circuits with identical speed tuning in
layer 4 are globally interconnected by divergence in the feedback
pathway. Excitatory and inhibitory inputs to the GRCs are interchanged for
feedback from cortical neurons with opposite direction preferences, as
indicated once for each type of feedback connection by the solid arrows.
Because of the long-range connections, each GRC receives modulatory input from
cortical cells jointly representing an extended visual field, with antagonism
between the two direction populations. Note that, rather than by divergence in
the corticogeniculate projection, the same function could alternatively be
implemented by lateral intracortical connections.}
\label{loop_net}
\end{figure}

As illustrated by Fig.\ \ref{vadapt_osegm}, cortical feedback signals that
carry information on local velocity measurements need to be sampled from an
extended visual field, spanning several RFs of cells in the primary visual
cortex. Lateral spread of information may be implemented either by
intracortical connections, or by divergence in the corticogeniculate feedback
pathway; cf.\ Fig.\ \ref{loop_net} right. Let us label all events of cortical
responses to local features within the entire population of layer 6 neurons
that feed back onto GRCs by an index $i$ according to their temporal order,
i.e., $t_1 < t_2 < \ldots$. Let furthermore ${\rm N}(t)$ be the number of such
events until time $t$. The slow dynamics of a GRC's membrane potential under
cortical control then is
\begin{equation}
V(t) \, = \, \sum_{i = 1}^{{\rm N}(t)} A_i \, {\rm PSP}(t - t_i) + V_0 \ ,
\label{V}
\end{equation}
where $A_i$ is the amplitude of the $i$th event of cortical feedback that
depends on the firing rate of a layer 6 neuron at time $t_i$. In this
formulation the size of the layer 6 population in the adaptive system, and
hence of the visual field from which motion information is integrated, is
represented by the number of response events per time: the rate of events
increases with the area from which motion is sampled.

Superimposed on the slow, cortically controlled dynamics [eqn.\ (\ref{V})] of
GRCs' membrane potentials is the relatively fast dynamics triggered by retinal
input. The retinal input and its consequences have been the subject of study
above. In the present model the dynamics of retinal input will not enter
explicitly. Instead, we now care only for its {\em cortical effect}, that is,
for cortical velocity tuning.

Cortical response rates in layer 4 to moving visual features such as bars are
assumed to follow some kind of velocity-tuning function. We have demonstrated
above that the velocity that produces the maximal convergent input from lagged
and nonlagged geniculate relay cells is, not surprisingly, close to the
velocity that yields {\em coincident} lagged and nonlagged response peaks,
corresponding to ${\rm t}_{\rm nl}(v) - {\rm t}_{\rm l}(v) \approx 0$; cf.\
Fig.\ \ref{fig3} third column. In this spirit we define a layer 4 cell's
response rate to a feature moving with velocity $v$ into the cell's preferred
direction as
\begin{equation}
R \, = \, {\rm f}\left[|{\rm t}_{\rm nl}(v) - {\rm t}_{\rm l}(v)|\right] \ ,
\end{equation}
where f is some positive, monotonically decreasing function. For analytical
treatment there will be some further restriction on f, to be formulated below.
Moreover, we assume layer 4 cells not to respond to features moving in their
null direction.

By looking at Fig.\ \ref{fig3} third column, we see that the simulation data
are roughly consistent with
\begin{equation}
\label{dt_fit}
\left[{\rm t}_{\rm nl} - {\rm t}_{\rm l}\right](v,V) \, = \,
{\rm s}(v) - {\rm c}(V) \ ,
\end{equation}
where s and c are monotonically decreasing functions of the stimulus velocity
$v$ and the relay cells' membrane potential $V$, respectively. This
approximation seems to be better at higher membrane potentials. Velocity tuning
is thus given by
\begin{equation}
R \, = \, {\rm f}\left[|{\rm s}(v) - {\rm c}(V)|\right] \ .
\label{R}
\end{equation}
Note that this family of functions is rather general in that it fits a large set
of velocity-tuning characteristics.

To obtain a physical interpretation of ${\rm s}(v)$ and ${\rm c}(V)$, we note
that the naive understanding of the velocity-dependence of ${\rm t}_{\rm nl}(v)
- {\rm t}_{\rm l}(v)$ is that it arises from the visual feature crossing the
lagged-nonlagged compound RF (see Fig.\ \ref{fig1}{\bf b}) in a time
proportional to $1/v$. Rewriting eqn.\ (\ref{dt_fit}) as
\begin{equation}
{\rm c}(V) \, = \, {\rm t}_{\rm l} - ({\rm t}_{\rm nl} - {\rm s}) \ ,
\end{equation}
it is evident that if s is the time it takes the visual feature to travel from
the lagged RFs to the nonlagged RFs, c is the difference in response times
intrinsic to the lagged and nonlagged populations. Appealing to this
interpretation, we will write
\begin{equation}
s_i := {\rm s}(v_i) \ , \quad {\rm RTD}(t) := {\rm c}[V(t)] \ , \quad
R_i := {\rm f}\left[|s_i - {\rm RTD}(t_i)|\right] \ ,
\end{equation}
and refer to $s_i$ as the {\em stimulus passage time} and to ${\rm RTD}(t)$ as
the (dynamic) {\em response time difference} between the lagged and nonlagged
populations. As above, the index $i$ enumerates the cortical response events in
temporal order. The response of a layer 4 neuron is thus maximal, if the
stimulus passage time equals the lagged-nonlagged response time difference.

We will use the simplest approximation for the dependence of the response time
difference on the membrane potential,
\begin{equation}
{\rm RTD}(t) \, = \, \chi V(t) + {\rm RTD}_0 \ ,
\label{RTD}
\end{equation}
that is, a linear function of $V$ with slope $\chi$ and offset ${\rm RTD}_0$.

In order to close the corticogeniculate loop, and the system of equations, we
need a transformation of the layer 4 responses $R_i$ to the feedback signals of
layer 6 cells, which in turn determine the amplitudes of postsynaptic
potentials $A_i$ in GRCs. To this end, we have analyzed a simple implementation
of cortical control. The rationale is that each cortical response $R_i$ to a
stimulus $s_i$ triggers a postsynaptic potential of amplitude $A_i$ through
feedback, either directly (depolarization, $A_i > 0$), or via the PGN or
geniculate interneurons (hyperpolarization, $A_i < 0$), such that the response
time difference ${\rm RTD}(t)$ is pulled {\em closer} to the stimulus passage
time $s_i$.

For the computation of the feedback signal a layer 4 cell's response
activity serves as a measure of the amount of mismatch between ${\rm RTD}(t_i)$
and $s_i$: low activity indicates large mismatch, high activity signals a good
match [cf.\ eqn.\ (\ref{R})]. The sign of mismatch is computed by comparing 
the outputs of layer 4 cells preferring the same direction but different
speeds, corresponding to different values of $V_0$ in eqn.\ (\ref{V}). A
simple network of spike-response neurons, shown in Fig.\ \ref{loop_net} left,
approximates these principles and provides the corresponding feedback signals
to the LGN. More precisely, computer simulations of this network show that
\begin{equation}
A_i \, \approx \, \gamma \, {\rm A}(R_i) \, \sigma(R^>_i - R^<_i) \ ,
\label{A}
\end{equation}
where A is some positive, monotonically decreasing function, $\sigma$ is a
sigmoidal function running between $-1$ and $+1$, and $R^>_i$, $R^<_i$ are
response rates of layer 4 cells tuned to higher and lower speeds than the cell
producing the response $R_i$. The factor $\gamma$ describes the overall
strength of cortical feedback to the LGN. Eqn.\ (\ref{A}) is the relation
we used between layer 4 responses and amplitudes of GRC potentials. Note that
it is the result of a simple transformation of the layer 4 activity to the
feedback signals of layer 6 cells; cf.\ Fig.\ \ref{loop_net} left. Although
a LGN-layer-4-layer-6-LGN loop of synaptic connections is indeed supported
by anatomical data \cite{Katz87,Sherman&Guillery96}, our aim was primarily
to investigate a network as simple as possible that can do the necessary
computation.

For the above control mechanism to work, sets of cortical layer 4 neurons are
required that have overlapping RFs and {\em differ} in speed tuning. It is not
necessary, but conceptually most straightforward, to assume the same set of $n$
classes of neurons, defined by the initial values $V_0^{(c)}$ [cf.\ eqn.\
(\ref{V})], $c = 1,2,\ldots,n$, of their membrane potentials, to represent each
retinal location. Furthermore, we want to restrict our attention to cortical
neurons that prefer one out of two opposite directions of motion. For each
class $c$ of cortical neurons there are two variants selective for the two
opposite directions, to be labeled by the superscripts $(c,+)$ and $(c,-)$. The
interaction between the ``$+$'' and the ``$-$'' population is taken to be such
that features moving in opposite directions elicit PSP amplitudes $A_i$ of
opposite signs in each GRC. This kind of antagonism ensures that the average
adaptation to incoherent, directionally unbiased motion is zero and, hence, is
vital to object segmentation; cf.\ Fig.\ \ref{vadapt_osegm}. We will label the
direction of a moving object by ``$+$'' and the opposite direction by ``$-$''.

Counting stimulus passage times $s_i$ as positive in the ``$+$'' and negative
in the ``$-$'' direction, the complete system dynamics is described by the equations
\begin{eqnarray}
\label{V^(c)}
V^{(c,\pm)}(t) & = & \pm \sum_{i = 1}^{{\rm N}(t)} ({\rm sgn} s_i) A_i^{(c)} \,
{\rm PSP}(t - t_i) + V_0^{(c)} \ ,\\
\label{R^(c)}
R_i^{(c)} & = & {\rm f}\left[|s_i - {\rm RTD}^{(c,{\rm sgn} s_i)}(t_i)|\right] \ ,\\
\label{RTD^(c)}
{\rm RTD}^{(c,\pm)}(t) & = & \pm \left[\chi V^{(c,\pm)}(t) +
{\rm RTD}_0\right] \ ,\\
\label{A^(c)}
A_i^{(c)} & = & \gamma \, {\rm A}(R_i^{(c)}) \, \sigma(R_i^{(c + 1)} - R_i^{(c - 1)}) \ .
\end{eqnarray}
In the last equation one may define $c_0 \equiv c_1$ and $c_{n+1} \equiv c_n$
to deal with the lowest and highest values of the class index. The stimulus
passage times $s_i$ and the times $t_i$ of the cortical responses are
stochastic variables and depend on the statistics of the stimulus. Adaptation
of velocity tuning and the dynamics of cortical responses are thus described as
a stochastic process driven by the stimulus. Figure \ref{loop_net} attempts to
give a picture of the complete system of corticothalamic loops.

The precise shapes of the functions f, A, and $\sigma$ are unimportant, as long as
the combination ${\rm A} \circ {\rm f}$ can be approximated reasonably well by
a linear function within a relevant range of values. A full analytical treatment is
possible for the limiting cases
\begin{equation}
{\rm A}\left[{\rm f}(|x|)\right] \, = \, |x| \ ,\quad
\sigma(x) \, = \, {\rm sgn}(x) \ .
\end{equation}
Computer simulations were run using
\begin{equation}
{\rm A}\left[{\rm f}(|x|)\right] \, = \,
\left\{ \begin{array}{ll} |x| & \mbox{if $|x| < p$,} \\
p & \mbox{elsewhere.} \end{array} \right. \ ,\quad
\sigma(x) \, = \, \tanh(q \, x) \ ,
\end{equation}
with positive parameters $p$ and $q$.

\subsection*{Results for the corticogeniculate loop}

A detailed mathematical analysis will be published elsewhere. Here we describe
the main results in the form of prose and present computer simulations for
illustration.

\begin{figure}
\centerline{\includegraphics[width=0.6\textwidth]{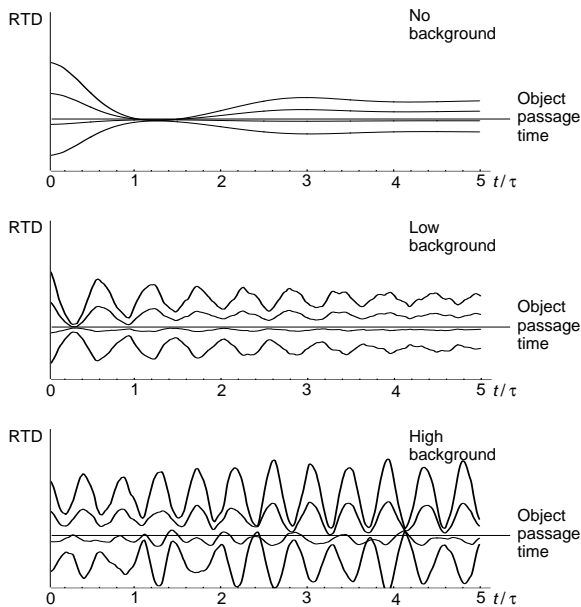}}
\caption[]{Computer simulations of the adaptation dynamics for the response
time differences ${\rm RTD}^{(c,\pm)}(t)$ of 4 classes of neurons ($c =
1,2,3,4$). Their receptive fields are arranged in an array of $60 \times 60$
retinal locations with 8 neurons per location. The time scale is set by $\tau$,
the rise time of corticothalamic postsynaptic potentials; cf.\ eqn.\
(\ref{PSP}). The stimulus is a coherently moving object (15 dots) without a
background (top), with a moderate background (10 dots, middle), and with a
strong background (30 dots, bottom). Background dots move at various random
speeds, exceeding or less than the object's speed, in or opposite to the
direction of object motion; cf.\ Fig.\ \ref{vadapt_osegm}. Only the 4
trajectories ${\rm RTD}^{(c,+)}(t)$, i.e., representative of the neurons
selective for the direction of object motion, are shown. The passage time of
object dots is indicated by the straight horizontal line in each graph. The
adaptation dynamics converges to a stationary state close to the object's
passage time without a background, and is increasingly oscillatory with
increasing background strength. The feedback strength is $\gamma = 0.03$ for
the no-background case, and $\gamma = 0.25$ for the two other cases.}
\label{3_adapt_tracks}
\end{figure}

When the system is stimulated by a coherently  moving object on a background of
incoherent, directionally unbiased  motion, the layer 4 cells' preferences
approach the velocity {\em of the object}. This results in an enhanced
representation of a coherent object and suppression of background features
moving at different velocities, in accordance with the scheme formulated in Fig.\
\ref{vadapt_osegm}. More specifically, with an absent or just a sparse
background, the response time differences ${\rm RTD}^{(c,+)}(t)$, i.e.,
the preferences of neurons encoding motion in the object's direction, settle in a
{\em stationary} state close to the object's passage time; see Fig.\
\ref{3_adapt_tracks} top. As background is added to the stimulus, ongoing
oscillations of the neurons' preferences develop. These oscillations are more
pronounced with a stronger background; see Fig.\ \ref{3_adapt_tracks} middle
and bottom. In other words, with a dense background, the dynamics is  {\em
oscillatory} and alternating between phases of close and poor matching of 
velocity preferences to the object's velocity; see Fig.\ \ref{3_adapt_tracks}
bottom.  We suggest to call the latter phenomenon a {\em diffusion-sustained
oscillation}, emphasizing the diffusion-like drive away from a stationary state
that is due to background features.

The response time differences ${\rm RTD}^{(c,\pm)}(t)$ of all layer 4 neurons
in the system are correlated, both across space and class $c$. The correlation
across space is the result of the lateral connectivity of the system; cf.\
Fig.\ \ref{loop_net} right. The correlation across classes is mediated by the
stimulus; cf.\ Fig.\ \ref{3_adapt_tracks}. In particular, neurons of all
classes $(c,+)$ are well tuned to an object's speed at the {\em same} time. At
such times, therefore, object features elicit strong responses in neurons of
{\em all} classes $(c,+)$, yielding a high {\em class-averaged} response,
\begin{equation}
\label{<R>c}
\smean{R^+} \, := \, \frac{1}{n} \sum_{c=1}^n R^{(c,+)} \ ;
\end{equation}
cf.\ eqn.\ (\ref{R^(c)}). At other times the preferences of neurons not only
are untuned with respect to the object, but also vary significantly across
neurons of different classes $c$. Hence there are no strong class-averaged
responses $\smean{R^+}$, neither to an object, nor to background features.

\begin{figure}
\centerline{\includegraphics[width=\textwidth]{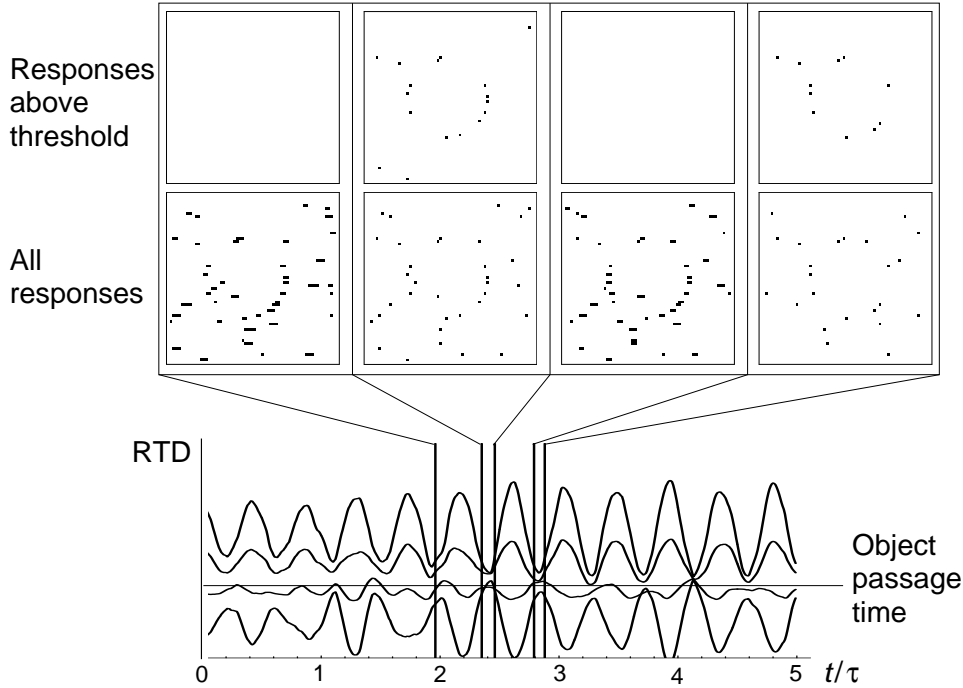}}
\caption[]{Spatiotemporal pattern of cortical activity in the $60 \times 60$
retinotopic array of neurons associated with the adaptation dynamics depicted
at the bottom of Fig.\ \ref{3_adapt_tracks}, re-displayed at the bottom of this
figure. The system is stimulated by 15 dots placed on a circle, all moving at a
{\em common} velocity to the right (object), and 30 dots moving at  various
random speeds, exceeding or less than the object's speed, either to the left or
to the right; cf.\ Fig.\ \ref{vadapt_osegm}. Whenever the neurons' preferences,
i.e., response time differences (RTDs), get {\em close to} the object's passage
time (corresponding to the object's speed), object dots elicit a strong
response in neurons that represent their actual retinal positions. The
corresponding array of activity is shown in the panels at the top of the
figure. The lower row of panels displays {\em all} responses that occur within
the indicated intervals of time; the upper row displays only those responses
where the class-averaged activity $\smean{R^+}$ [see eqn.\ (\ref{<R>c})] for a
location {\em exceeds} a certain threshold. Nearly all supra-threshold activity
is related to object features, demonstrating the segmentation  of the object
against the background. Spurious supra-threshold activity derives from
responses to background dots that by chance have speeds very close to the
object's speed.  The supra-threshold activity is oscillatory and synchronous.
If $\tau$, the rise time of corticothalamic postsynaptic potentials [cf.\ eqn.\
(\ref{PSP})], is taken to be 100 ms, the oscillation frequency is roughly 25 Hz
in this example.}
\label{cortical_activity}
\end{figure}

In turns out that each time course of adaptation is associated with a certain
spatiotemporal pattern of cortical activity. For an adaptation time course such
as shown at the top of Fig.\ \ref{3_adapt_tracks}, there are permanent strong
class-averaged responses $\smean{R^+}$ to object features as soon as the RTDs
have converged to the object's passage time. On the other hand, an oscillatory
time course as shown at the bottom of the figure is associated with alternating
phases of weak and strong responses $\smean{R^+}$, the strong responses being
restricted to object features. In fact, the dynamic RTDs act as a {\em
pacemaker} for distributed cortical activity. A periodic time structure is
imposed that tends to synchronize the firing of layer 4 cells representing the
object; see Fig.\ \ref{cortical_activity}. Oscillations may show up in cortical
single cell activity or in multi-unit activity, depending on the nature of the
stimulus and the size of RFs. In particular, single neurons show periodic
activity, if their responses last longer than one period of oscillation of the
RTDs. Their response then comprises several ``elementary'' response events of
the kind marked by a single dot in Fig.\ \ref{cortical_activity}. The period
scales with the rise time $\tau$ of the corticothalamic postsynaptic potentials
[see eqn.\ (\ref{PSP})], and decreases to zero as the stimulus velocity or
density becomes very large. Hence, the period can be {\em much shorter} than
the duration of individual corticothalamic PSPs and even than their rise time
$\tau$. The reason is that for a stimulus consisting of features moving at
different velocities, PSPs are not initiated in isolation but superimposed upon
other PSPs of opposite signs.

It is worth noting that a stationary stimulus background and neuronal noise are
special cases of directionally unbiased background activity. The system thus
also performs segmentation of an object moving across a stationary background
and suppression of neuronal noise.

In the present context of stochastic dynamical systems, stability of the
dynamics means that at least the first two moments (the mean and the variance)
of the RTDs converge to finite values. It can be proved that stability is
guaranteed, if $\gamma$, the overall strength of cortical feedback to the LGN
[cf.\ eqn.\ (\ref{A})], is not too large.

To summarize, in this model corticogeniculate loops give rise to object
segmentation based on global sampling of velocity features in the visual field.
Visual features acquire an enhanced representation in the adaptive motion
system by virtue of belonging to a {\em coherent object}. The temporal
structure of adaptation and, hence, of cortical activity, depends on the
statistics of stimulus and noise. With oscillatory dynamics, different objects
detected by populations of neurons tuned to different directions of motion turn
out to be represented through statistically independent phases of oscillation;
cf.\ Gray et al.\ (1989)\nocite{Gray_etal89}. The model can be extended so as
to include higher visual areas and in such a way that phases of oscillation
become statistically independent for different two-dimensional (2D) directions
of motion, i.e., the true directions not accessible to single area 17 cells.

As a consequence of the described dynamics, the control loop for speed
preferences recruits neuronal hardware for object representation in a very {\em
economic} manner in that it avoids at the same time neurons that remain idle
and neurons that respond to irrelevant features.

\section*{Discussion}

Recently it has been proposed that corticogeniculate feedback modulates the
spatial layout of simple-cell RFs by exploiting the thalamic burst-tonic
transition of relay modes \cite{Worgotter_etal98}. Along a similar line, the
main point made by the first part of our modeling is that one should expect a
modulatory influence of cortical feedback on the spatio{\em temporal} RF
structure of simple cells. More precisely, we observe a shift in the time to
the bar-response peak that is {\em opposite} for lagged and nonlagged cells;
cf.\ Fig.\ \ref{fig2}. Assuming (i) an RF layout as usually found for
direction-selective simple cells in V1, and (ii) an influence of {\em
convergent} geniculate lagged and nonlagged inputs on this RF structure, it
follows that the observed shifts in response timing affect cortical speed
tuning. To the best of our knowledge, nobody has looked for such an effect yet.

Accepting the general conclusion on adaptive cortical speed tuning, the model
of the corticogeniculate loop demonstrates what such an effect in the brain may
be good for. We have attacked the fundamental sensory task of object
segmentation and analyzed a simple implementation of a control loop based on
abundant structures of the visual system. The basic idea is to select an
important, object-related component of motion. The loop mechanism effectively
draws samples of moving features from a retinal region and adapts to the mean
of those velocities that are more frequent in one direction than in the
opposite, that is, it adapts to an {\em asymmetric} peak in the allover
velocity density. Such peaks will usually be produced by objects moving over a
background; cf.\ Fig.\ \ref{vadapt_osegm}. Features moving at the corresponding
velocity acquire an enhanced representation.

The proposed adaptation mechanism could be characterized as {\em
pre}-atten\-tive, since it requires no ``act of will'' and no information from
higher brain areas where different objects are categorized. Multiple adaptive
motion systems could act in parallel for different directions of motion, in
different parts of the visual field, and on different spatial scales. From the
perspective of efficiency, higher perceptual and attentive mechanisms can be
envisaged best to operate on top of a representation level where
object-unrelated activity is reduced, as provided by the adaptive motion
system. Although at this point it is difficult to pin down the precise
mechanisms of contextual effects on neuronal responses in the visual system, it
seems clear that modulation related to object segmentation occurs as early as
in V1, at least in primates; see Lamme et al.\ (1998)\nocite{Lamme_etal98} and
references therein.

A critical experimental test of a control loop such as the one analyzed here is
to look for changes in spatiotemporal RF structure on a timescale of, say, 100
ms in response to moving, especially large-field, stimuli. It would be
interesting to study reverse correlations in the speed or temporal-frequency
domain, analogous to work that has uncovered dynamic orientation tuning in the
monkey \cite{Ringach_etal97}.

More detailed modeling is required to compare the oscillations and
synchronization that arise in the loop model under certain conditions with
recent data on persistent oscillations in area 17 that are preferentially
evoked by {\em moving} stimuli \cite{Bringuier_etal97,Castelo-Branco_etal98}.
In the loop model, oscillations may arise from visual stimulation {\em at}
geniculocortical synapses, showing up as periodic excitatory synaptic
potentials in cortical neurons. Because of the slow potentials that modulate
the GRC states we would expect such oscillations to be at the lower end of the
observed frequency spectrum (below, say, 30 Hz); cf.\ Fig.\
\ref{cortical_activity}. Clearly, since such periodic inputs are not restricted
to layer 4 \cite{Bringuier_etal97}, intracortical connections have also a role
to play, certainly in the propagation of the periodic activity.

As expounded above, the loop model was not designed to fit any particular set
of experimental data, and at this stage it does not explain convincingly
cortical oscillations. Whatever the details of a control loop, however, a
slow modulation of GRC membrane potentials controlled by comparatively fast
responses of cortical neurons is likely to produce oscillations under certain
conditions. Our argument thus does add a new dimension to the search for the
origin of rhythmic and synchronous activity in the low-frequency range.

Great care must be taken when extrapolating from cats to primates. No lagged
relay cells have been described in the primate LGN so far. On the other hand, a
recent study \cite{DeValois&Cottaris98} does suggest a set of geniculate inputs
to directionally selective simple cells in macaque striate cortex that is
essentially analogous to the lagged-nonlagged set envisaged for cat simple
cells. If indeed primates employ a similar system at the lowest level of motion
analysis, the corticogeniculate loop would implement at this level the
classical Gestalt principle of ``common fate'', which recognizes common motion
as a powerful cue for object segmentation in humans
\cite{Wertheimer58,Julesz71}. Furthermore, we would obtain a new rationale to
the nature of the input to neurons in primate area V2 that extract the shape of
objects from coherent motion on a background
\cite{Peterhans&Baumann94,Peterhans97}. That is to say, V1 {\em detects} an
object through common motion and V2 then {\em analyzes} its shape.

An important side effect of the adaptation mechanism is that it can account for
visual stabilization during fixational eye movements. These erratic eye
movements cause retinal slips providing motion signals that are directionally
{\em unbiased} on the timescale of the control loop ($\sim$ 100 ms). Since
unbiased motion does not lead to adaptation to any particular velocity, none of
the moving features is highlighted. Now, if the brain relies on the activity of
adaptively speed-tuned neurons to infer motion in the outside world, no such
motion will be inferred from fixational eye movements. Stabilization of the
visual world is thus achieved because successive motion signals cancel out in
the control loop.

Interestingly, this stabilization mechanism explains the recently reported
`jitter after-effect' \cite{Murakami&Cavanagh98}. An observer is first exposed
to dynamic random noise for 30 s, and then fixates a larger area of static
random noise. The perception is a coherent jitter of the static noise that
follows fixational eye movements, specifically, in the region that has {\em
not} been exposed to the prior random motion. The explanation in terms of the
present model relies on the fact that responses of velocity-selective neurons
are reduced after stimulation for 30 s [also called `adaptation', but unrelated
to the adaptation of speed preferences proposed here; see Hammond et al.\
(1988)\nocite{Hammond_etal88}]. As can be seen from eqn.\ (\ref{A}), for {\em
reduced} cortical responses $R_i$, $R^>_i$, $R^<_i$ to each of the sweeps
induced by eye movements, the amplitudes $A_i$ of the corticogeniculate
potentials [cf.\ eqn.\ (\ref{V})] become {\em larger}\footnote{More precisely,
eqn.\ (\ref{A}) tells us that the amplitudes $A_i \rightarrow 0$ as the cortical
responses grow to their maximum or decrease to zero. Hence, maximal amplitudes
$A_i$ are produced in between.}. The effect is similar to an increased allover
strength $\gamma$ of cortical feedback. So it may be that each brief period of
eye-induced motion that normally averages out in adaptation now gives rise to
fluctuations of adaptation of sufficient amplitude to evoke {\em enhanced}
responses outside the region of reduced responsiveness. Thus the same mechanism
that ensures stabilization during normal fixation will now signal jitter
motion.

The results presented here offer new explanations for data from diverse
research. They also suggest new experimental paradigms to test their
implications. In case of confirmation, a re-interpretation of a large amount of
data on motion processing might be needed.

\section*{Acknowledgments}

The authors thank Esther Peterhans for stimulating discussions concerning her
experimental data, and Christof Koch and Mark H\"ubener for valuable comments
on the manuscript. U.H.\ was supported by the Deutsche Forschungsgemeinschaft
(DFG), grant GRK 267.

\end{document}